\documentclass[a4paper,11pt]{article}
\usepackage{pos}

\usepackage{amsfonts}
\usepackage{amsmath}
\usepackage{cancel}
\usepackage{graphicx}
\usepackage{hyperref}
\usepackage{multirow}
\usepackage{makecell}
\usepackage{xspace}
\usepackage{placeins}
\usepackage{nicefrac}

\definecolor{colorDG}{HTML}{008000}

\newcommand{\rn}{{\tt r}}
\newcommand{\Mn}{{\tt M}}
\newcommand{\R}{\text{Re}}
\newcommand{\I}{\text{Im}}

\newcolumntype{P}[1]{>{\centering\arraybackslash}p{#1}}

\title{Global fit of the Aligned Two-Higgs-Doublet Model}

\author*[a]{Anirban Karan}
\emailAdd{kanirban@ific.uv.es}

\author[b,c]{V\'ictor Miralles,}
\emailAdd{victor.miralles@manchester.ac.uk}

\author[a]{Antonio Pich}
\emailAdd{antonio.pich@ific.uv.es}

\affiliation[a]{Instituto de F\'isica Corpuscular, Parque Cient\'ifico, Catedr\'atico Jos\'e Beltr\'an 2, E-46980 Paterna, Spain}

\affiliation[b]{INFN, Sezione di Roma, Piazzale A. Moro 2, I-00185 Roma, Italy}

\affiliation[c]{Department of Physics and Astronomy, University of Manchester, Oxford Road, Manchester M13 9PL, United
Kingdom}

\abstract{Though the Standard Model (SM) provides a very elegant description of the interactions among fundamental particles, there are ample evidences suggesting that new physics is needed.
In particular, extending the scalar sector has enough motivation from vacuum stability, electroweak phase transition and various other sectors. Among different such extensions, the two-Higgs-doublet model (THDM) is the simplest one that preserves the electroweak $\rho$ parameter. 
Flavour-changing neutral currents (FCNC) are usually avoided 
by implementing additional discrete symmetries, 
but this type of models are subject to
severe phenomenological constraints. 
In the more general framework of the aligned THDM (ATHDM) tree-level FCNCs are avoided by choosing the same flavour structure for the Yukawa couplings of the two scalar doublets, 
which results in weaker phenomenological constraints.
Here, we present a global fit of the ATHDM, using the package HEPfit that performs a bayesian analysis on the parameter-space of this model with the help of stability and perturbativity bounds, experimental data for various flavour and electroweak precision observables, and constraints from Higgs searches at the LHC. This global fit has been performed assuming that all additional scalars are heavier than the SM Higgs and that there are no extra sources of CP violation beyond the CKM phase.
}

\FullConference{The European Physical Society Conference on High Energy Physics (EPS-HEP2023)\\
 21-25 August 2023\\
Hamburg, Germany\\}

\begin{document}
\maketitle

\section{The Aligned Two-Higgs-Doublet Model}

The THDM \cite{Branco:2011iw} is one of the simplest extensions of the SM where, in addition to all the SM particles, we have another scalar doublet with $Y=1/2$. In the ``Higgs basis'' only one of the two doublets acquires a vacuum expectation value $v=246$ GeV. One charged ($G^\pm$) and one neutral ($G^0$) scalars act as Goldstone bosons providing masses to the $W^\pm$ and $Z$ bosons. The physical scalar spectrum contains one charged and three neutral bosons leading to a very rich phenomenology. 
In the CP-conserving scenario, the two CP-even neutral scalars $(S_{1,2})$ mix together to create the mass eigenstates $h$ and $H$, leaving the CP-odd scalar $S_3$ unmixed:
{\small
\begin{equation}
\Phi_1=\frac{1}{\sqrt 2}\begin{pmatrix}
\sqrt 2\;G^+\\
S_1+v+i\, G^0
\end{pmatrix}\, ,\;\;\Phi_2=\frac{1}{\sqrt 2}\begin{pmatrix}
\sqrt 2\;H^+\\
S_2+i\, S_3
\end{pmatrix}
\;\longrightarrow\;
\begin{pmatrix}
h\\H
\end{pmatrix}=\begin{pmatrix}
\cos\tilde{\alpha}& \sin\tilde{\alpha}\\ -\sin\tilde{\alpha}&\cos\tilde{\alpha}
\end{pmatrix}\begin{pmatrix}
S_1\\S_2
\end{pmatrix}\;\text{and}\; A=S_3\, .
\end{equation}
}

The scalar potential of this model can be expressed as
{\small\begin{align}
\label{eq:pot}
\mathcal V&=\mu_1\,\Phi_1^\dagger \Phi_1+\mu_2\,\Phi_2^\dagger \Phi_2+ \Big[\mu_3\,\Phi_1^\dagger \Phi_2+h.c.\Big]+\frac{\lambda_1}{2}\,(\Phi_1^\dagger \Phi_1)^2+\frac{\lambda_2}{2}\,(\Phi_2^\dagger \Phi_2)^2+\lambda_3\,(\Phi_1^\dagger \Phi_1)(\Phi_2^\dagger \Phi_2)\nonumber\\
&+\lambda_4\,(\Phi_1^\dagger \Phi_2)(\Phi_2^\dagger \Phi_1)+\Big[\Big(\frac{\lambda_5}{2}\,\Phi_1^\dagger \Phi_2+\lambda_6 \,\Phi_1^\dagger \Phi_1 +\lambda_7 \,\Phi_2^\dagger \Phi_2\Big)(\Phi_1^\dagger \Phi_2)+ \mathrm{h.c.}\Big]\, ,
\end{align}}
where all the parameters are real in the CP-conserving case. The minimization condition relates $\mu_{1}$ and $\mu_{3}$ to $\lambda_1$ and $\lambda_6$: $v^2=-2\mu_1/\lambda_1=-2\mu_3/\lambda_6$. Thus the number of independent parameters in the scalar sector becomes nine: $\mu_2\, , v\, , \lambda_{1,..,7}$. Since it is more convenient to work with the physical masses and mixing angle, we use to following relations:
\begin{align}
\label{eq:mix_ang_lam}
&\tan\tilde\alpha=\frac{M_h^2 -v^2\,\lambda_1}{v^2\,\lambda_6}=\frac{v^2\,\lambda_6}{v^2\,\lambda_1-M_H^2}\, , \quad M_{h,H}^2=\frac{1}{2}\,(\Sigma\mp\Delta),\quad M_A^2=M_{H^\pm}^2+\frac{v^2}{2}\,(\lambda_4-\lambda_5)\, , \nonumber\\
&M_{H^\pm}^2=\mu_2+\frac{\lambda_3}{2}\,v^2\,  \text{ with } \;\Sigma=M_{H^\pm}^2+\Big(\lambda_1+\frac{\lambda_4}{2}+\frac{\lambda_5}{2}\Big)\,v^2\; \text{ and }\; \Delta=\sqrt{\big(\Sigma-2\lambda_1 v^2\big)^2+4\,\lambda_6^2\,v^4}\, ,
\end{align}
to choose our nine independent parameters to be: $v,\,M_{H^\pm},\,M_{h},\,M_{H},\,M_{A},\,\tilde{\alpha},\,\lambda_2,\,\lambda_3$ and $\lambda_7$. On the other hand, 
the couplings of the neutral scalars with the gauge bosons can be expressed as: $
g_{hVV}=\cos\tilde{\alpha}\;g_{hVV}^{SM}\, , \; g_{HVV}=-\sin\tilde{\alpha}\;g_{hVV}^{SM}\, \text{and} \; g_{AVV}=0$ \
($VV\equiv W^+W^-, ZZ$).

In the basis of fermion mass eigenstates, the 
Yukawa Lagrangian takes the form ($\tilde\Phi_a\equiv i \tau_2 \Phi^*_a$):  
\begin{equation}
    -\mathcal L_Y=\big(\sqrt 2\big/{v}\big)\;\Big\{\bar Q_L (M_u \tilde\Phi_1 + Y_u \tilde\Phi_2)u_R + \bar Q_L (M_d \Phi_1 + Y_d \Phi_2)d_R + \bar L_L (M_\ell \Phi_1 + Y_\ell \Phi_2)\ell_R + \mathrm{h.c.} \Big \} \, ,
\end{equation}
where $M_f$ $(f\equiv u,d,\ell)$ are the diagonal mass matrices of the fermions and $Y_f$ are arbitrary $3\times3$ 
matrices generating tree-level FCNCs. However, imposing $Y_f=\varsigma_f \, M_f$ (with real $\varsigma_f$ in the CP-conserving case) the tree-level FCNCs can be avoided and the Yukawa Lagrangian becomes:
{\small\begin{equation}
-\mathcal L_Y=\sum\Big({y_f^{\varphi^0_i}}\big/{v}\Big)\,\varphi^0_i\,\Big[\bar f M_f \mathcal{P}_R f\Big]+\big(\sqrt 2\big/{v}\big)\, H^+\,\Big[\bar u\,\big\{\varsigma_d V M_d \mathcal{P}_R-\varsigma_u M_u^\dagger V\mathcal{P}_L\big\}\, d+\varsigma_\ell\, \bar \nu M_\ell \mathcal P_R \ell\Big] + \mathrm{h.c.}\, ,
\end{equation}}
where $\mathcal P_{L,R}=(1\mp\gamma^5)/2$\,, $\varphi_i^0\,\equiv \{h,H,A\}$, $V$ is the CKM matrix and the Yukawa couplings are:
\begin{align}
&y_{f}^H=-\sin\tilde\alpha+\varsigma_{f}\,\cos \tilde\alpha\, , \qquad y_{f}^h=\cos\tilde\alpha+\varsigma_{f}\,\sin \tilde\alpha\, , \qquad
y_{u}^A= - i\varsigma_{u}\, , \qquad y_{d,\ell}^A=  i\varsigma_{d,\ell}\, .
\label{eq:Higgs_yuk_up}
\end{align}
 It is interesting to mention that the usual $\mathcal{Z}_2$ symmetric THDMs can be retrieved by imposing $\mu_3=\lambda_6=\lambda_7=0$ together with the following conditions: 
\begin{align}
&\text{Type I:\;\;} \varsigma_{u}=\varsigma_d=\varsigma_\ell=\cot\beta,\quad \text{Type II:\;\;} \varsigma_{u}=-\varsigma_d^{-1}=-\varsigma_\ell^{-1}=\cot\beta\, ,\quad 
\text{Inert:\;\;}\varsigma_{u}=\varsigma_d=\varsigma_\ell=0\, ,
\nonumber\\
&\text{Type X:\;\;} \varsigma_{u}=\varsigma_d=-\varsigma_\ell^{-1}=\cot\beta 
\qquad \text{and}\qquad
\text{Type Y:\;\;} \varsigma_{u}=-\varsigma_d^{-1}=\varsigma_\ell=\cot\beta \, .
\end{align}

\section{Fit setup}
For the global fit of the ATHDM, we use the open-source package HEPfit \cite{DeBlas:2019ehy} that works within a Bayesian statistics framework. Compared to the SM, here we have ten more parameters to fit and the priors chosen for the analysis are listed in Tab. \ref{tab:priors}. We have performed two fits, assuming that all additional scalars are heavier than the SM Higgs and varying their masses up to 1 TeV and 1.5 TeV, respectively. A linear prior was imposed for the variation of the masses. The range of $\tilde\alpha$ is chosen in such a way that the $5\sigma$ region of posterior probability lies within the range. The alignment parameters $\varsigma_f$ are varied taking into account the perturbativity of the Yukawa couplings, i.e. $\sqrt{2}\,|\varsigma_f| m_f/v<1$.
\begin{table}[htb]
{\renewcommand{\arraystretch}{0.9}
\resizebox{\textwidth}{!}{%
\small
\begin{tabular}{|P{1.1cm}|P{1.1cm}|P{1.1cm}|P{1.1cm}|P{1.1cm}|P{1.1cm}|P{1.1cm}|P{1.1cm}|P{1.1cm}|P{1.1cm}|P{1.1cm}|P{1.1cm}|}
\hline
\multicolumn{12}{|c|}{Priors} \\
\hline
\hline
\multicolumn{4}{|c}{$M_{H^\pm} \subset$ [0.125, 1.0\, (1.5)] TeV} & \multicolumn{4}{|c|}{$M_{H} \subset$  [0.125, 1.0\, (1.5)] TeV} & \multicolumn{4}{c|}{$M_{A} \subset$  [0.125, 1.0\, (1.5)] TeV} \\
\hline
\multicolumn{4}{|c}{$\lambda_2 \subset$ [0, 11]} & \multicolumn{4}{|c|}{$\lambda_3 \subset$ [-3, 17]}  & \multicolumn{4}{c|}{$\lambda_7 \subset$ [-5, 5]}  \\
\hline
\multicolumn{3}{|P{3.3 cm}}{$\tilde{\alpha} \subset$ [-0.16, 0.16]} & \multicolumn{3}{|P{3.3 cm}|}{$\varsigma_u \subset$ [-1.5, 1.5]} & \multicolumn{3}{P{3.3 cm}|}{$\varsigma_d \subset$ [-50, 50]} & \multicolumn{3}{P{3.3 cm}|}{$\varsigma_\ell \subset$ [-100, 100]} \\  
\hline
\end{tabular}
}
}
\caption{Priors chosen for the new-physics parameters.}
\label{tab:priors}
\end{table}

\section{Theoretical constraints}
\begin{figure}[h!]
    \centering
    \includegraphics[scale=0.3]{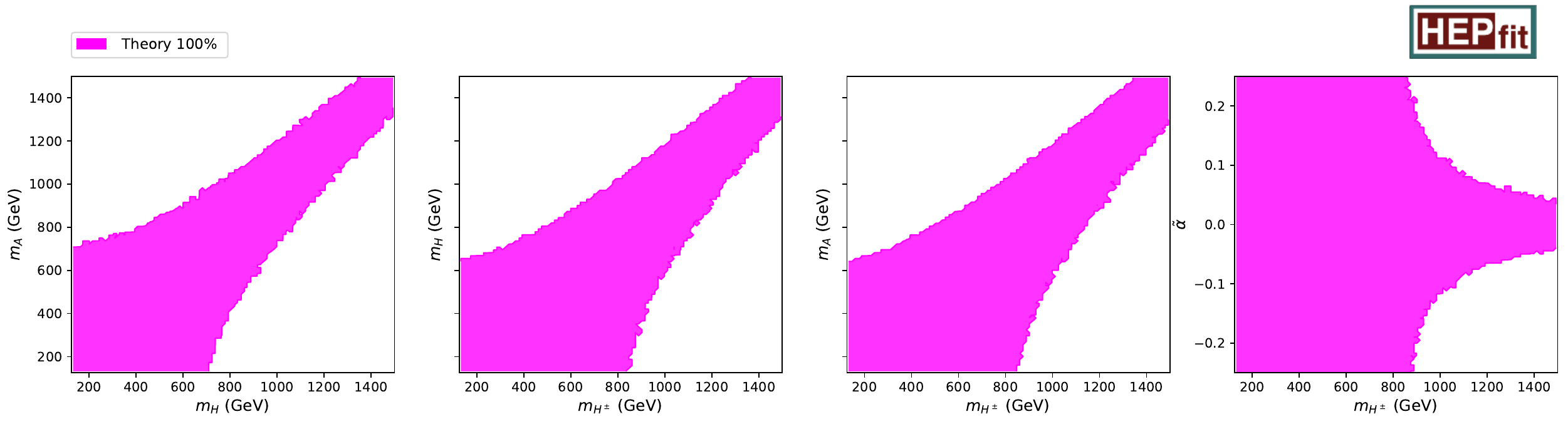}
\includegraphics[scale=0.3, trim={0in 0.0in 0in 0.75in},clip]{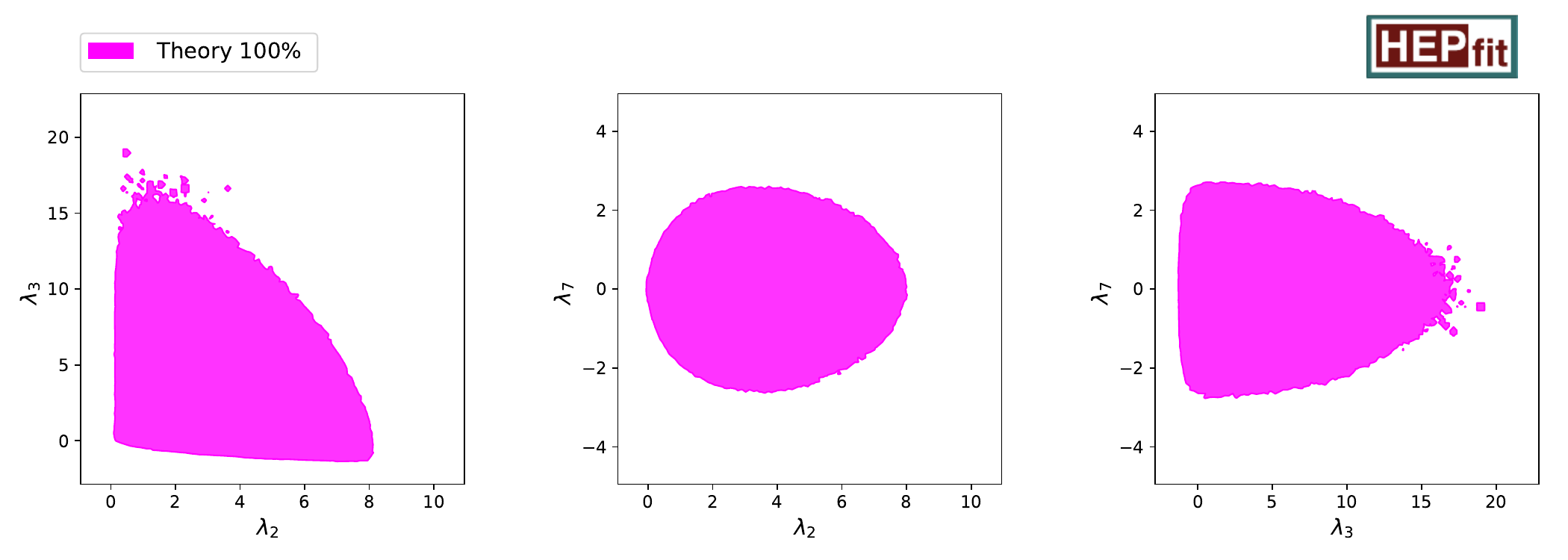} 
    \caption{Theoretical constraints on the parameter-space of the ATHDM.}
    \label{fig:theory}
\end{figure}
There are two constraints on the model from the theoretical side: a) the scalar potential should be bounded from below, and b) perturbative unitarity must hold for the S-matrix. 

To impose the ``bounded from below'' condition, one first constructs the
Minkowskian 4-vector
$\rn^\mu\, =\, \Big(|\Phi_1|^2+|\Phi_2|^2, \, 2\,\R (\Phi_1^\dagger\Phi_2), \, 2\,\I (\Phi_1^\dagger\Phi_2), \, |\Phi_1|^2-|\Phi_2|^2\Big)$, and writes down the scalar potential as $\mathcal V=-\,\Mn_\mu\,{\rn}^\mu + \nicefrac{1}{2}\,\Lambda^{\mu}_{\phantom{\mu}\nu}\, \rn_\mu\,\rn^\nu$. After diagonalization of the mixed-symmetric matrix $\Lambda^{\mu}_{\phantom{\mu}\nu}$\,, the ``bounded from below'' condition of the scalar potential is ensured if \cite{Ivanov:2015nea}: 1) all the eigenvalues of $\Lambda^{\mu}_{\phantom{\mu}\nu}$ are real, and 2) the ``timelike'' eigenvalue $\Lambda_0$  is larger than the three ``spacelike'' eigenvalues $\Lambda_{1,2,3}$, along with $\Lambda_0>0$. Moreover, the vacuum can be guaranteed to be a stable neutral minimum by imposing $D>0$, or $D<0$ with $\xi>\Lambda_0$, where $D=\mathrm{Det}[\xi\, {\mathbb I_4}-\Lambda^{\mu}_{\phantom{\mu}\nu}]
$ and $\xi=\big(m_{H^\pm}^2\big/v^2\big)$.

To guarantee perturbative unitarity one first constructs
the matrix of tree-level partial-wave amplitudes 
for all the $2\to2$ scatterings involving scalars and Goldstones 
and then demands the eigenvalues of the S-wave amplitudes at very high energy to satisfy $(a_0^{0})^2\leq \frac{1}{4}$, where
$(\mathbf{a_0})_{i,f}=\frac{1}{16\pi s}\int_{-s}^{0} dt \;\mathcal M_{i\to f}(s,t)$.
%
The theoretical constraints restrict the masses and quartic couplings of the scalars to a great extent and these restrictions with 100\% probability are depicted in Fig. \ref{fig:theory}.

\section{Electroweak precision observables}
\begin{figure}[h!]
    \centering
    \includegraphics[scale=0.35]{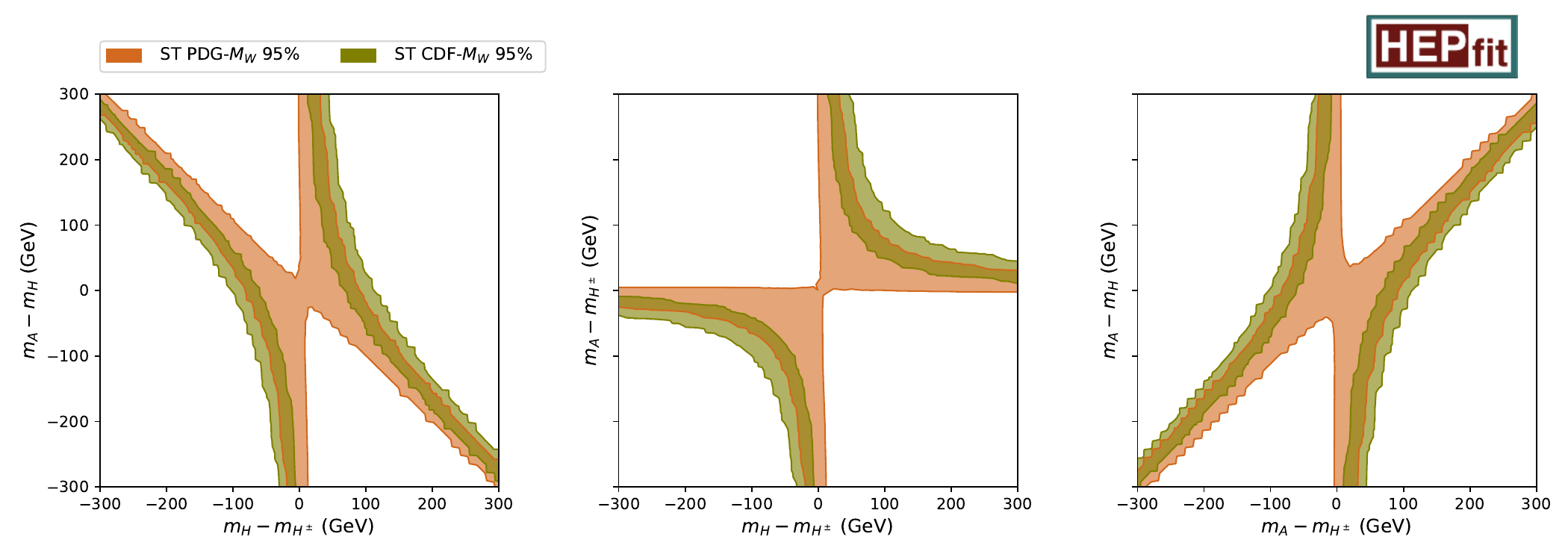}
    \caption{Constraints from the oblique parameters S and T.}
    \label{fig:EWPO}
\end{figure}

The presence of additional scalars alters the values of the oblique parameters S, T, U. We first perform the electroweak fit, removing $R_b\equiv \Gamma(Z\rightarrow b\bar{b})/\Gamma(Z\rightarrow \rm{hadrons})$ that also acquires contributions from extra scalars, and then use S and T (THDM contributions to U are suppressed) for our fits. Since the oblique parameters are sensitive to $M_W$, we perform two different fits using the PDG \cite{ParticleDataGroup:2022pth} and CDF \cite{CDF:2022hxs} values of $M_W$ which are shown in Fig. \ref{fig:EWPO}. While the PDG value of $M_W$ is compatible with zero mass splitting among the scalars, the CDF value does not allow it.

\section{Flavour constraints}
\begin{figure}[h!]
    \centering
\includegraphics[scale=0.318]{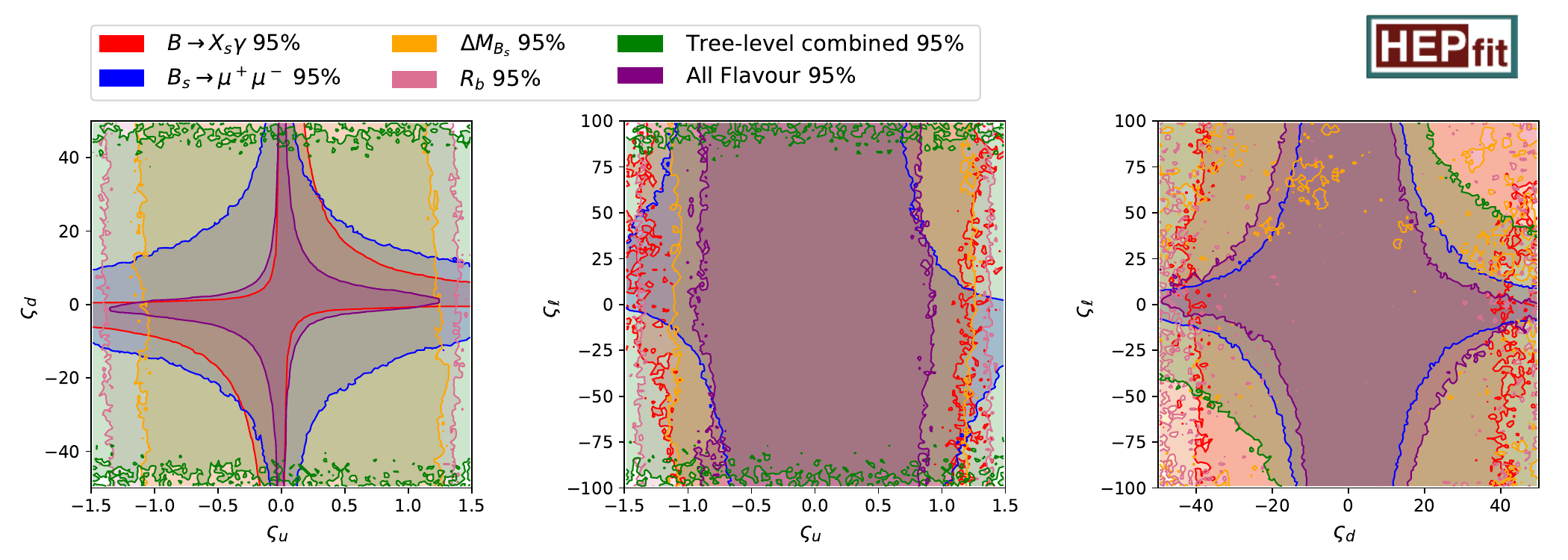}\hfill
\includegraphics[height=3.75cm, width=3.75cm, trim={0in 0in 5.5in 0.0in},clip]{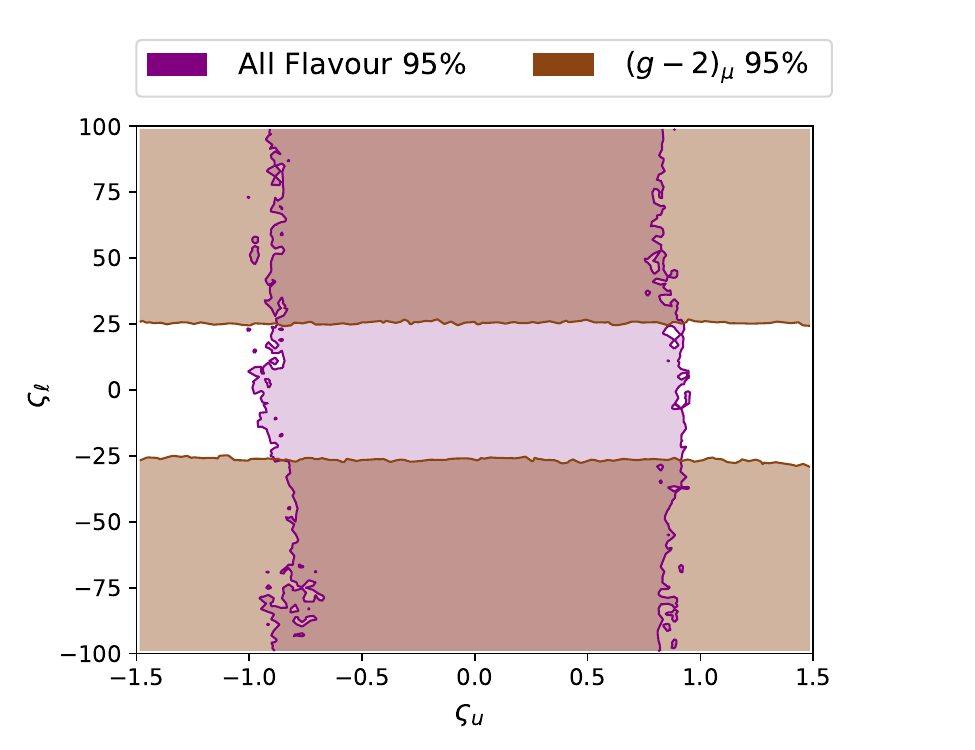}
    \caption{Constraints from flavour observables.}
    \label{fig:Flavour}
\end{figure}

In the flavour sector, we consider the constraints from loop-induced processes like $B_s^0-\bar B_s^0$ mixing ($\Delta M_{B_s}$), $B\to X_s \gamma$ and $B_s\to \mu^+\mu^-$, relevant tree-level transitions like $B\rightarrow \tau \nu$, $D_{(s)}\rightarrow \mu \nu$ and $D_{(s)}\rightarrow \tau \nu$, as well as ratios of leptonic  decay widths of light pseudoscalar mesons like $\Gamma(K\rightarrow \mu\nu)/\Gamma(\pi\rightarrow \mu\nu)$ and ($\Gamma(\tau\rightarrow K\nu)/\Gamma(\tau\rightarrow \pi\nu)$), along with $R_b$. These observables mainly restrict the alignment parameters, which are shown in Fig. \ref{fig:Flavour}. It is important to mention that we have fitted the CKM parameters separately, using only processes which are not contaminated by the additional scalars. Fig. \ref{fig:Flavour} compares also the fitted value of $\varsigma_\ell$, obtained from the combination of all flavour observables, with the value of $\varsigma_\ell$ required for explaining the muon $(g-2)$.

\section{Higgs signal strengths and direct searches}

\begin{figure}[h!]
    \centering
    \includegraphics[scale=0.35]{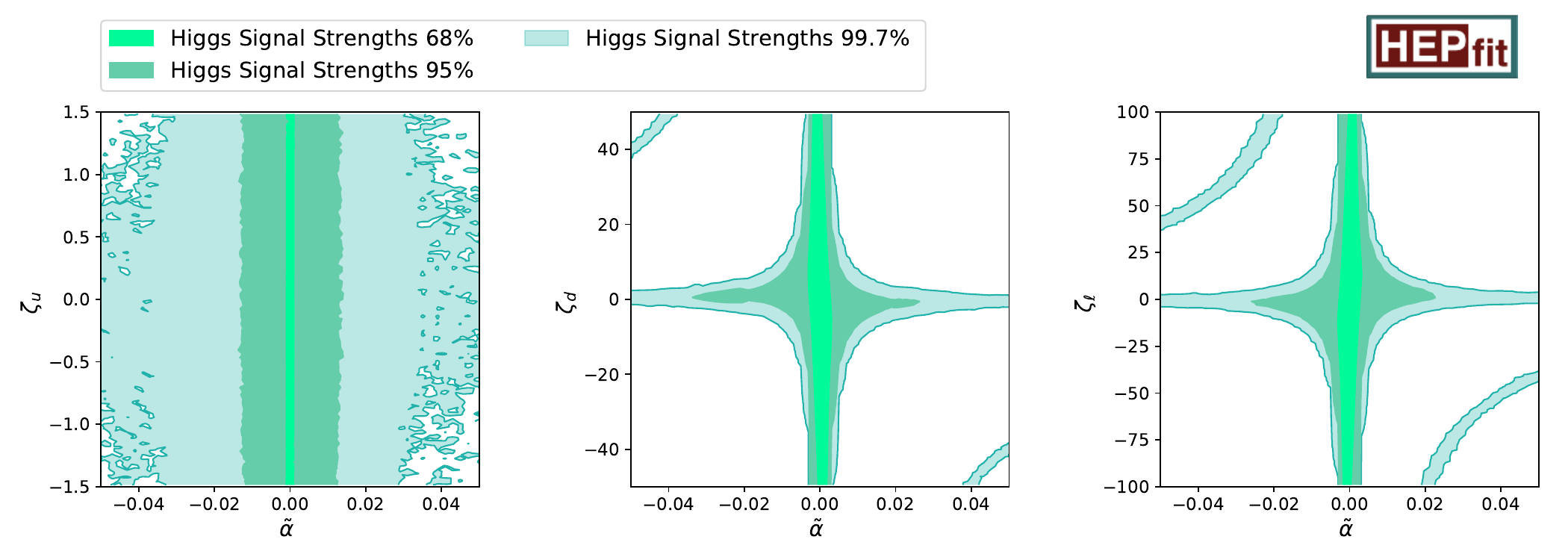}
    \caption{Bounds from Higgs signal strengths.}
    \label{fig:sig_str}
\end{figure}

Production and subsequent decay of the SM Higgs have been measured at the LHC through the production modes ggF, VBF, Vh and tth, and the decay channels to $c\bar{c}$, $b\bar{b}$, $\gamma\gamma$, $\mu^+\mu^-$, $\tau^+\tau^-$, $WW$, $Z\gamma$ and $ZZ$. 
As shown in Fig.~\ref{fig:sig_str}, these data put stringent bounds on the mixing angle $\tilde\alpha$ and the alignment parameters $\varsigma_{d,\ell}$,
while the constraints on $\varsigma_u$ are weaker than the corresponding limits from flavour data.
Additionally, one can also observe the wrong-sign solutions for the Yukawa couplings $y^h_d$ and $y^h_\ell$ at the corner regions of the second and third plots in Fig. \ref{fig:sig_str}.

We have also compared the cross section times branching fraction for different processes with the exclusions limits from ATLAS and CMS. All the data on Higgs signal strengths and direct searches that have been included in the fit are listed in Ref. \cite{Karan:2023kyj}.

\section{Global fit}
A summary of the marginalised probabilities obtained from the global fit is given in Tab.~\ref{tab:marginalised_results}, where the limits of our two baseline fits are shown. There is a small dependence on the priors adopted, since allowing higher masses favours higher values of the Yukawa alignment parameters (the heavier the scalars the less they contribute to flavour and direct searches for the same value of the Yukawa alignment parameter) while disfavours larger values of the mixing angle (the heavier the scalars the smaller the mixing angle must be to fulfill the theory assumptions). The correlations among observables (and a more detailed discussion) are shown in Ref.~\cite{Karan:2023kyj}.

\label{sec:globa_fits_results}

\begin{table}[h!]
{\renewcommand{\arraystretch}{0.9}
\resizebox{\textwidth}{!}{%
\small
\begin{tabular}{|P{1.cm}|P{1.cm}|P{1.cm}|P{1.cm}|P{1.cm}|P{1.cm}|P{1.cm}|P{1.cm}|P{1.cm}|P{1.cm}|P{1.cm}|P{1.cm}|}
\hline
\multicolumn{12}{|c|}{\bf Marginalised Individual results} \\
\hline
\hline
\multicolumn{12}{|c|}{ \it Masses up to 1 TeV} \\
\hline
\multicolumn{4}{|P{4. cm}}{$M_{H^\pm} \ge$ 390 GeV} & \multicolumn{4}{|P{4. cm}|}{$M_{H} \ge$ 410 GeV} & \multicolumn{4}{P{4. cm}|}{$M_{A} \ge$ 370 GeV} \\
\hline
\multicolumn{4}{|P{4. cm}}{$\lambda_2$: $3.2 \pm 1.9$} & \multicolumn{4}{|P{4. cm}|}{$\lambda_3$: $5.9 \pm 3.5$}  & \multicolumn{4}{P{4. cm}|}{$\lambda_7:$ $0.0 \pm 1.1$}  \\
\hline
\multicolumn{3}{|P{3.8 cm}}{$\tilde{\alpha}:$ $(0.05 \pm 21.0)\cdot10^{-3}$} & \multicolumn{3}{|P{3. cm}|}{$\varsigma_u:$ $0.006 \pm 0.257$} & \multicolumn{3}{P{3. cm}|}{$\varsigma_d:$ $0.12 \pm 4.12$} & \multicolumn{3}{P{3.3 cm}|}{$\varsigma_\ell:$ $-0.39 \pm 11.69$} \\  
\hline
\hline
\multicolumn{12}{|c|}{ \it Masses up to 1.5 TeV} \\
\hline
\multicolumn{4}{|P{4.4 cm}}{$M_{H^\pm} \ge$ 480 GeV} & \multicolumn{4}{|P{4.4 cm}|}{$M_{H} \ge$ 490 GeV} & \multicolumn{4}{P{4.4 cm}|}{$M_{A} \ge$ 480 GeV} \\
\hline
\multicolumn{4}{|P{4.4 cm}}{$\lambda_2$: $3.2 \pm 1.9$} & \multicolumn{4}{|P{4.4 cm}|}{$\lambda_3$: $5.9 \pm 3.8$}  & \multicolumn{4}{P{4.4 cm}|}{$\lambda_7:$ $0.0 \pm 1.2$}  \\
\hline
\multicolumn{3}{|P{3.8 cm}}{$\tilde{\alpha}:$ $(0.8 \pm 16.8)\cdot10^{-3}$} & \multicolumn{3}{|P{3.3 cm}|}{$\varsigma_u:$ $-0.011 \pm 0.407$} & \multicolumn{3}{P{3.3 cm}|}{$\varsigma_d:$ $-0.096 \pm 6.22$} & \multicolumn{3}{P{3.3 cm}|}{$\varsigma_\ell:$ $-1.18 \pm 17.54$} \\  
\hline
\end{tabular}
}
}
\caption{Global fit results. The mass limits are at 95\% probability while for the others we show the mean value and the square root of the variance. }
\label{tab:marginalised_results}
\end{table}

\section{Conclusion}


We have performed a global fit of the ATHDM, including both theoretical constraints and experimental data. We have taken into account the available experimental information, updating the results of Ref.~\cite{Eberhardt:2020dat} with the most recent data.
The regions preferred by the data have some dependence on the adopted priors,  since the scalar masses cannot be bounded from above. When the masses are allowed to vary up to 1 TeV, values of $|\varsigma_u|>0.7$ , $|\varsigma_d|>12$,  $|\varsigma_\ell|>30$ and  $|\tilde{\alpha}|>0.06$ lay outside the 95\% probability region.

\section*{Acknowledgements}

This work has been supported by Generalitat Valenciana (grant PROMETEO/2021/071), MCIN/AEI/10.13039/501100011033 (grant No. PID2020-114473GB-I00), the Italian Ministry of Research (MUR, grant PRIN20172LNEEZ) and the ERC under the European Union’s Horizon 2020 research and innovation programme (Grant agreement No. 949451).


\end{document}